\documentclass[aps,prl,showpacs,twocolumn,preprintnumbers,amsmath,amssymb]{revtex4-1}

\usepackage{graphicx}
\usepackage{graphicx}
\usepackage{dcolumn}
\usepackage{bm}
\usepackage{color}
\usepackage{soul}

\newcommand{\ST}[1]{\textcolor{black}{#1}}
\usepackage[breaklinks=true]{hyperref}
\newcommand{\red}{\textcolor{black}}

\begin{document}
\title{Phase transitions of photon fluid flows driven by a virtual all-optical piston}
\author{Abdelkrim Bendahmane$^{1}$}
\email{contributed equally}
\author{Gang Xu $^{1}$}
\email{contributed equally; current address: Department of Physics, The University of Auckland, Private Bag 92019, Auckland 1142, New Zealand}
\author{Matteo Conforti$^{1}$ }
\author{Alexandre Kudlinski$^{1}$ }
\author{Arnaud Mussot$^{1,2}$ }
\author{Stefano Trillo$^{3}$}
\email{stefano.trillo@unife.it} 
\affiliation{$^{1}$ Univ. Lille, CNRS, UMR 8523 - PhLAM - Physique des Lasers Atomes et Mol{\'e}cules, F-59000 Lille, France}
\affiliation{ $^{2}$Institut Universitaire de France (IUF)}
\affiliation{$^{3}$Department of Engineering, University of Ferrara, Via Saragat 1, 44122 Ferrara, Italy}
\affiliation{}

\date{\today} 
 
\begin{abstract} 
The piston problem, i.e. the dynamics in a uniform gas at rest under the action of a moving piston is fundamental problem of physics and a \ST{canonical case study} in shock wave physics. We investigate theoretically and experimentally the analogous problem for a photon fluid, which turns out to be strongly influenced by the dispersive character of the problem. The experiment makes use of a fiber optics setup where an all-optically controlled quasi instantaneous change of frequency of the input light mimics the piston action. We show that the flow exhibits phase transitions which involve cross-over from regimes characterized by 2-shocks (pushing piston) to 2-rarefaction waves (retracting piston), with the appearance of cavitating states at critical amplitudes of the jump. Importantly, the appearance of vacuum points into the 2-shock marks the transition to a regime which is unique to the photon fluid. Our observations allow for the extensive and quantitative test of the state of the art description of the dispersive Riemann problem, i.e. Whitham modulation theory, applied to the universal defocusing nonlinear Schr\"odinger equation.
\end{abstract} 
 

 
\maketitle 

\section{\label{sec:intro} Introduction}
Dispersive hydrodynamics, i.e. the fluid-like flows in media which exhibit dispersive (or diffractive) effects but no relevant viscous effects is attracting a growing interest aimed at highlighting the common traits of physical phenomena of different origin \cite{Whitham_book,ElHoefer16}.
Undoubtedly, the most generic and impressive manifestation of the dispersive nature of the hydrodynamic-like flows is the appearance of unsteady fast coherent oscillatory wavetrains known as Dispersive Shock Waves (DSWs). Contrary to classical hydrodynamics, where the shock waves are often associated to irreversible phenomena and 
viscous effects rule the actual thickness of the non-undulatory shock front, DSWs provide the generic mechanism of dispersive regularization of a gradient catastrophe in reversible systems. Spectacular manifestations of such process occurs in natural phenomena (strong tidal bores going upstream along river estuaries \cite{Chanson_book12} or atmospheric gravity waves \cite{atmospheric}), and has been sporadically reproduced in the lab in the past  \cite{Taylor70,HS74part2,Rothenberg89}. However, it is only recently that DSWs has been systematically observed and understood under reproducible experimental conditions in areas as different as bulk \cite{Wan07,Ghofraniha07,Conti09,Wang15} or fiber \cite{Handbook,Fatome14,Xu16,Wetzel16,Millot1617,Xu17,Nuno19} nonlinear optics, superfluid Helium \cite{Rolley07helium}, dilute Bose-Einstein condensates (BEC) \cite{Dutton01,Hoefer06}, electron waves \cite{Mo13}, viscous fluid conduits \cite{Maiden16} and water waves in flumes \cite{Trillo16}, spin waves in magnetic films \cite{Janantha17}, and also investigated in disordered media \cite{Ghofraniha12}, incoherent waves \cite{Garnier13,Xu15ncomm}, and in connection to radiative phenomena \red{\cite{Conforti13,Conforti14,Malaguti14,Smyth16}.}

Importantly, the optical experiments are described by the defocusing nonlinear Schr\"odinger equation (NLSE), which allows to conceive the propagating light as a {\em photon fluid} that behaves in close analogy to superfluid quantum many-body systems with repulsive interaction (see also \cite{Carusotto14,Vocke15}). Remarkably, this analogy stands on the purely conservative ground (unlike other photon fluids such as polaritons in semiconductor cavities \cite{polariton}), with optical power and phase derivative (chirp) playing the role of density and velocity in the hydrodynamical formulation of the NLSE or  Gross-Pitaevskii mean-field model.

On one hand, considering the light as a photon fluid allows to probe new regimes of dispersive hydrodynamics, that might be difficult to observe in pure quantum fluids. On the other hand, this poses the challenging problem of assessing the analogies or potential disparities with the behavior of classical fluids such as ideal gases and liquids. In this respect, it is crucial to investigate in photon fluids the canonical problems of shock wave propagation in gas dynamics (or well known equivalent in water waves \cite{Gilmore50}). A recent step forward in this direction is the observation of the decay of an initial step in the density, which realises a particular configuration of the Riemann problem, equivalent to the well known gas shock tube problem (i.e., the flow determined by an initial jump in density) in gas dynamics or dam-breaking problem in water  \cite{Xu17,Janantha17,Janosi04}. In this case, the photonic dam decays into a coexisting rarefaction wave (RW) and DSW connected by a plateau, with a threshold for DSW cavitation (i.e., the appearance of null point in the DSW) occurring in remarkable agreement with predictions of modulation theory (or Whitham averaging \cite{ElHoefer16,Hoefer06,Pavlov87,Bikbaev95,El95,Kodama99}) applied to the defocusing NLSE \cite{focusing}.

\begin{figure}[htb]
\centering
\includegraphics[width=8.3cm]{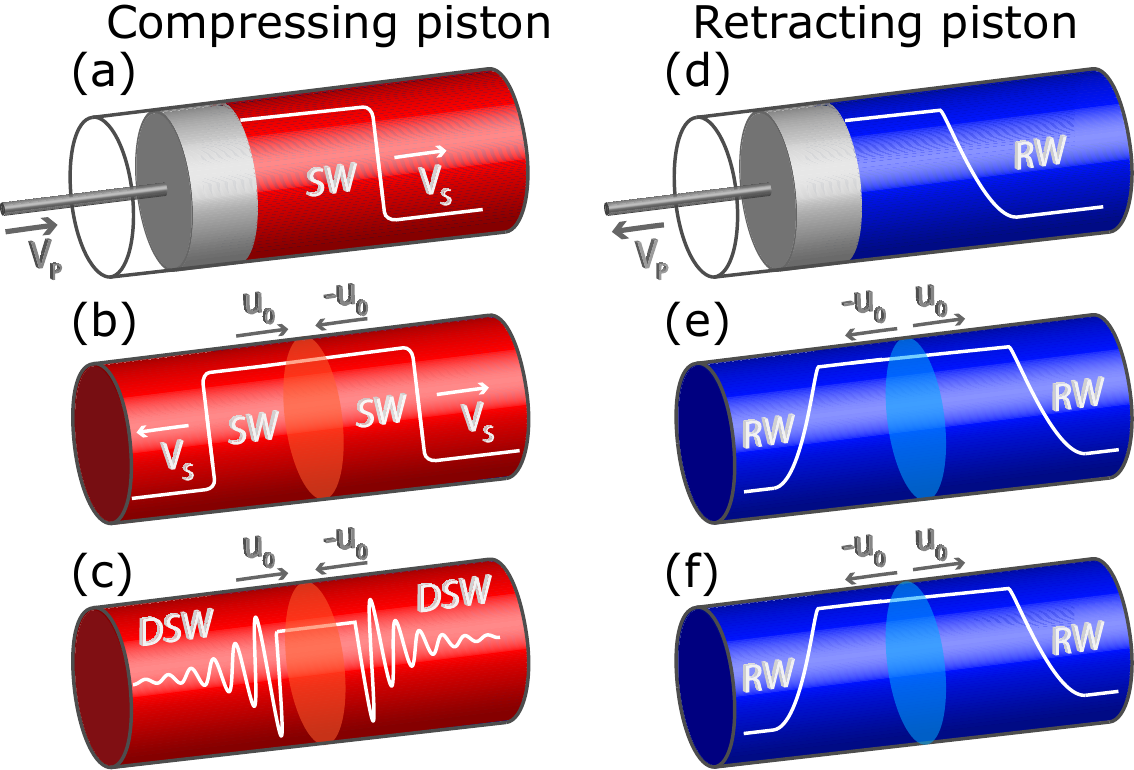}
     \caption{
Schematic of the physics (white curves show qualitatively the density) a,b,c: Shock wave dynamics for "pushing" piston:
(a) compressing piston producing a shock wave in an ideal (non-dispersive) fluid;
(b) 2-shock produced via step-wise initial velocity profile (Riemann problem);
(c) DSW-DSW via Riemann problem in a dispersive photon fluid;
d,e,f: Rarefaction wave dynamics for "retracting" pistons:
(d) single RW;
(e) RW-RW from the Riemann problem;
(f) similar 2-RW in a dispersive photon fluid.}
\label{fig:cartoon}
\end{figure}

In this paper, we address a different case study in the field of shock waves that arises from gas dynamics, namely the {\em piston problem}, i.e. the determination of the flow induced in an ideal uniform gas by a piston set impulsively into motion with constant velocity. In the framework of the studies of classical shock waves (CSWs) which lasted for more than a century \cite{Whitham_book,Courant_book,Griffith54,Krehl_book,JC98,Leveque_book}, such problem is a paradigm which has the canonical solution (see \cite{CoulsonJeffrey_book,Kevorkian_book} and Appendix A in \cite{Supp}) schematically illustrated in Fig. \ref{fig:cartoon}(a,d)\red{. When} the piston compresses the gas at rest on its right, a CSW emerges that travels ahead of the piston with supersonic velocity dictated by Rankine-Hugoniot condition \cite{Krehl15}, whereas a retracting piston produces a smooth RW. Our aim is to implement the analog problem for photons. To this end, we exploit its conceptual identity with a Riemann problem (evolution of step initial data  \cite{Toro_book}), where the physical piston is replaced by a suitably prepared initial condition characterized by a stepwise variation of fluid velocity over a constant density (virtual pistons).
In gases, the latter ideally produces a bi-directional replica of the CSW or RW, as sketched in Fig. \ref{fig:cartoon}(b,e). In particular, when the initial velocities are pointing inwards [Fig. \ref{fig:cartoon}(b)], two counter-propagating CSWs emerge, which can be understood as the density fronts associated to the velocity jumps from $u_0$ to zero and zero to $-u_0$, respectively. Conversely, initial velocities pointing outwards result into two RWs that expand in opposite directions [Fig. \ref{fig:cartoon}(e)]. In the dispersive regime characteristic of the photon fluid, the RWs remain unaltered due to their smoothness [Fig. \ref{fig:cartoon}(f)], whereas the CSWs are turned into expanding DSWs [Fig. \ref{fig:cartoon}(c)]. The formation of the two DSWs, however, is expected to exhibit critical behavior \cite{ElHoefer16,Bikbaev95,El95}.
Indeed above a critical amplitude of the velocity jump where the two DSWs start to cavitate, they become connected by a nonlinear periodic wave instead of a constant background (as it is always the case in gases). This marks a phase transition to a regime which is unique to the photon fluid. A similar effect has been predicted when the piston is schematised as a moving potential in the defocusing NLSE \cite{Hoefer08}. A recent experiment in BEC, however, was found to deviate qualitatively from this scenario, rather exhibiting, in the long term, signatures of non-undulatory shock waves of the viscous type \cite{Mossman18}.

We report here the first observation of such a dispersive hydrodynamic transition. In particular, by employing a full fiber set-up, and imprinting step-like equivalent velocities by means of an ultrafast change of frequency chirp, we are able to report a number of intriguing results: (i) a comprehensive study of the phase transitions that occur in the dispersive piston problem ruled by the defocusing NLSE; (ii) the first observation of the regime mentioned above and unique to the photon fluid, where a fully undulatory solution emerges, featuring two DSWs connected through an unmodulated periodic wave; (iii) the observation of asymmetric DSWs and their critical transition to the fully undulatory solution that follows from the most general Riemann problem involving a simultaneous jump both in density and velocity (power and chirp in our setting).

A further very important point is that our results allow for a quantitatively accurate comparison (e.g. in terms of edge velocities of the DSWs, as well as critical values for transitions) with the theoretical results obtained by means of Whitham modulation theory applied to the NLSE. To this end, essential features of our experiment are the nearly conservative regime (loss compensation), the use of step-like initial data, and the absence of polarization effects (scalar regime). The latter two features are indeed strict hypothesis under which the modulation theory can be developed.
In this sense, we believe that our experiment represents a substantial step forward with respect to other optical experiments where sinusoidal phase modulation was used to observe flat-top pulses denoted as platicons \cite{Varlot13}, or vectorial configurations used to observe ballistic DSWs \cite{Nuno19}. Our configuration also substantially differs from other experiments proposed in BEC based on piston-like action \cite{El09obstacle,Pinsker13}.

\section{\label{sec:th} Theory of Riemann problem}

We consider the following complex electric field $E(T,Z) \exp(i k_0 Z - i \omega_0 T_{lab})$, where $\omega_0=2\pi c/\lambda$ is the carrier frequency, $k_0$ the relative propagation constant, $Z$ is the distance along the fiber, $T_{lab}$ is the time in the rest frame, and the envelope $E(T,Z)$ obeys the NLSE
\begin{equation} \label{nls}
i  \frac{\partial E}{\partial Z} - \frac{k''}{2}\frac{\partial^{2} E}{\partial T^2} +  \gamma \left| E \right|^{2} E = 0,
\end{equation}
where $T=T_{lab}-k'_0 Z$ is the retarded time in a frame moving with group-velocity $1/k'_0$. We make use of a dispersion compensating fiber (DCF), operating at $\lambda_0=1561$ nm  with normal dispersion $k''=d^2k/d\omega^2 \vert_{\omega_0}=170$ ps$^2$/km and nonlinear coefficient $\gamma=3$ (W km)$^{-1}$. Since we operate in the {\em defocusing} regime of the NLSE ($\gamma k''>0$), a clear connection to CSWs of gas dynamics does exist. It becomes manifest by applying the Madelung transform $E(T,Z)=\sqrt{P_0} \sqrt{\rho(t,z)} \exp \left(-i \int_{-\infty}^{t} u(t',z) dt' \right)$, which allows to formulate the NLSE in fluid dynamics form:
\begin{eqnarray} 
& &\rho_z + (\rho u)_t = 0\;;\label{swe1} \\ 
& &u_z + \left( \frac{u^2}{2} + \rho \right)_t= \frac{1}{4}\left[ \frac{\rho_{tt}}{\rho} - \frac{(\rho_t)^2}{2\rho^2} \right]_t, \label{swe2}
\end{eqnarray}
where we set $z=Z/Z_0$, $t=T/T_0$ with $Z_0 \equiv (\gamma P_0)^{-1}$ and $T_0 \equiv \sqrt{k''/(\gamma P_0)}$, $P_0$ being a reference power. By neglecting the right hand side arising from dispersion and containing higher-order derivatives (also known as quantum pressure term \cite{Wan07}), Eqs. (\ref{swe1}-\ref{swe2}) are identical to the {\em dispersionless} vector Eulerian conservation law that rules the dynamics of the one-dimensional flow in an isentropic gas with pressure law $p \sim \rho^2$ \cite{Leveque_book}. Here, the normalized power $\rho(t,z)=|E(T,Z)|^2/P_0$ plays the role of local gas density, whereas the normalized chirp $u(t,z)=\Delta \omega(T,Z) T_0$ is equivalent to gas velocity. Here $\Delta \omega(T,Z)=-d \phi/dT$ is the dimensional chirp or local instantaneous frequency deviation, expressed in terms of the envelope phase $\phi=Arg[E(Z,T)]$. Note also that space and time have interchanged roles compared with gas dynamics. In this limit, which henceforth will be referred to as the {\em dispersionless} NLSE, the resulting system is hyperbolic,
thereby admitting weak solutions known as CSWs, which describe traveling jumps in density and velocity. 

\begin{figure}[htb]
\centering
\includegraphics[width=8cm]{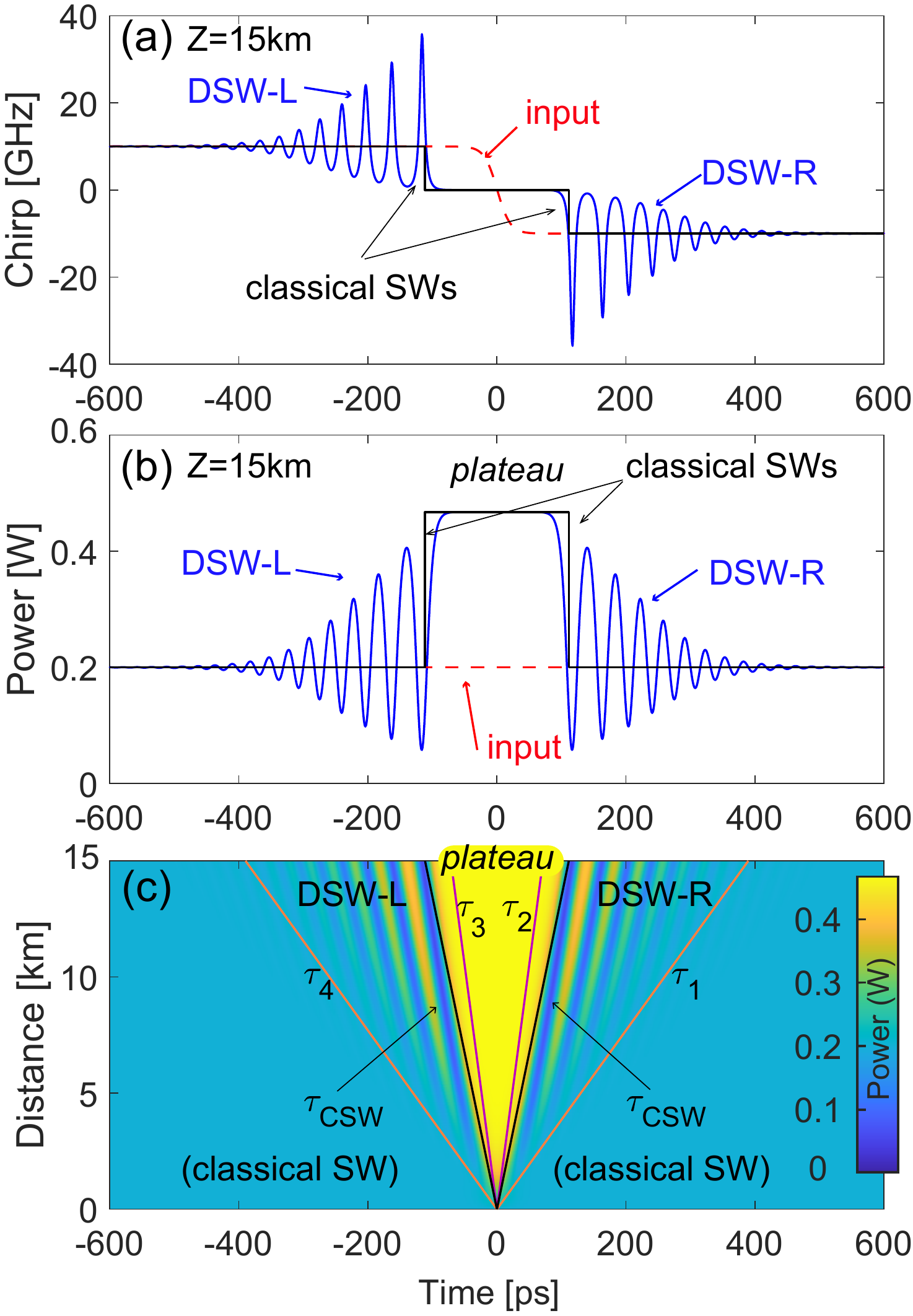}
     \caption{Formation of two DSWs ruled by the full NLSE contrasted with the CSWs of the dispersionless (isentropic gas-dynamic) case. 
     \red{(c)} Evolution of power $|E(Z,T)|^2= \rho(z,t) P_0$ in $T-Z$ plane showing the expansion fans of the two left (DSW-L) and right (DSW-R) shocks delimited by edge velocities in Eq. (\ref{Whitham2DSW});
     \red{(a,b)} Snapshot of power (b) and chirp $\Delta f(T,Z)=u(t,z)/(2 \pi T_0)$ \red{(a)} at the output $Z=15$ km (blue solid line). The input (dashed red) is a frequency step $\Delta f(z=0)=\pm 10$ GHz on a constant power $P_0=200$ mW. The CSWs are the solid black lines in \red{(a,b)}, while their velocities [Eq. (\ref{vel2CSW})] are given by the oblique black lines in \red{(c)}. The fiber parameters are those of the experiment.
    }
\label{fig:2shocks}
\end{figure}

In particular, the step-like initial condition shown in Fig. \ref{fig:cartoon}(b), i.e. a decreasing  jump in velocity (chirp) from $u(t<0,z=0)=u_0$ to $u(t>0,z=0)=-u_0$, with $u_0>0$, on top of a density $\rho=$ constant produces, in the dispersionless limit, two CSWs propagating in opposite directions as illustrated by the black solid lines in Fig. \ref{fig:2shocks}. These CSWs are fully equivalent to the shocks induced by two pistons pushing a gas in opposite directions (mathematical details of such equivalence are discussed in Appendix A in \cite{Supp}). As shown in Fig. \ref{fig:2shocks}(b,c), the CSWs feature two fronts that connect either the initial left ($\rho,u_0$) or the right ($\rho,-u_0$) quiescient state, to a plateau or intermediate state of constant density $\rho_i>\rho$ and zero velocity ($u_i=0$), which emerges spontaneously and turns out to be expanding as the two shocks propagate. The explicit expression of the intermediate density \begin{equation}\label{rhoi}
\rho_i=\left( \frac{u_0}{2} + \sqrt{\rho} \right)^2
\end{equation}
follows from a simple-wave approach for hyperbolic equations, which allows also to express the velocities of the shocks from the well-known Rankine-Hugoniot condition \cite{Leveque_book,Krehl15}.  In terms of the self-similar variable $\tau=t/z$ \cite{warnvelocity}, such velocities reads
\begin{equation} 
\tau_{CSW}^{\pm}=\mp \left(\frac{u_0}{4} - \sqrt{\rho} \right), \label{vel2CSW}
\end{equation}
where the upper (lower) sign refers to the right (left) CSW.
In Fig. \ref{fig:2shocks}, we also contrast this gas dynamics scenario with the corresponding dispersive dynamics obtained by numerical integration of the full NLSE. As shown, in the latter case, the two shocks become indeed dispersive, being characterized by expanding fans (see \red{Fig. \ref{fig:2shocks}(c)}) where fast oscillations spontaneously appear, connecting the upper and lower quiescient states. The wavetrains that constitute the left (DSW-L) and right (DSW-R) dispersive shocks reflect their nature of periodic nonlinear waves (dn-oidal waves) with strongly modulated parameters \cite{ElHoefer16} over an interval that ranges from a soliton edge (the inner deep end of the wavetrain) to an harmonic edge (where oscillations become shallower, i.e. quasi linear). These two edges travel with different velocities, say $\tau_{1,2}$ for the DSW-R, which can be predicted by means of Whitham modulation theory (details in Appendix B in \cite{Supp}) and reads as:
\begin{equation} 
\tau_{1}=\frac{u_0^2 + 3 u_0 \sqrt{\rho} +  \rho}{\sqrt{\rho} + u_0};  ~~\tau_{2}=\sqrt{\rho}-\frac{u_0}{2}. \label{Whitham2DSW}
\end{equation}
It is also clear from Fig. \red{\ref{fig:2shocks}(c)} that the edge velocities of the DSW-L are $\tau_{4}=-\tau_{1}$ and $\tau_{3}=-\tau_{2}$ due to symmetry,
whereas the CSW velocities are bounded within the DSW fans.

The case illustrated above is actually a particular case of the most general Riemann problem such that the initial condition is step-like in both chirp and power.
In real world units, the classification of the dynamics depends on four arbitrary parameters, namely the boundary values across the chirp and power jumps. 
However, using normalized variables $\rho$ and $u$, without loss of generality, we can assume, initial conditions characterised by only two parameters $(\rho_0, u_0)$:
\begin{eqnarray} \label{inival}
\rho(t,0)=1+(\rho_0 -1) \theta(t);~ u(t,0)=u_0-2u_0  \theta(t),
\end{eqnarray}
where $\theta(t)=[1+{\rm sign}(t)]/2$ is the Heaviside unit step function. Equation (\ref{inival}) implies a left to right symmetric jump in velocity, from $u_L=u_0$ to $u_R=-u_0$, accompanied by a power jump from $\rho=\rho_L=1$ to $\rho=\rho_R=\rho_0$ \cite{stepnormalization}. This allows us to classify all the possible evolution scenarios in a simple parameter plane $(\rho_0, u_0)$. According to the general theory of $2 \times 2$ conservation laws \cite{Leveque_book}, the decay of the step initial data ruled by the {\em dispersionless} NLSE can occur through the generation of a pair of fundamental waves, each being of the shock or rarefaction type, separated by a constant state. Therefore, three possible combinations can emerge: (i) CSW-CSW, (ii) CSW-RW, (iii) RW-RW, depending on the value of the initial data $(\rho_0, u_0)$.

\begin{figure*}[ht]
\includegraphics[width=18cm]{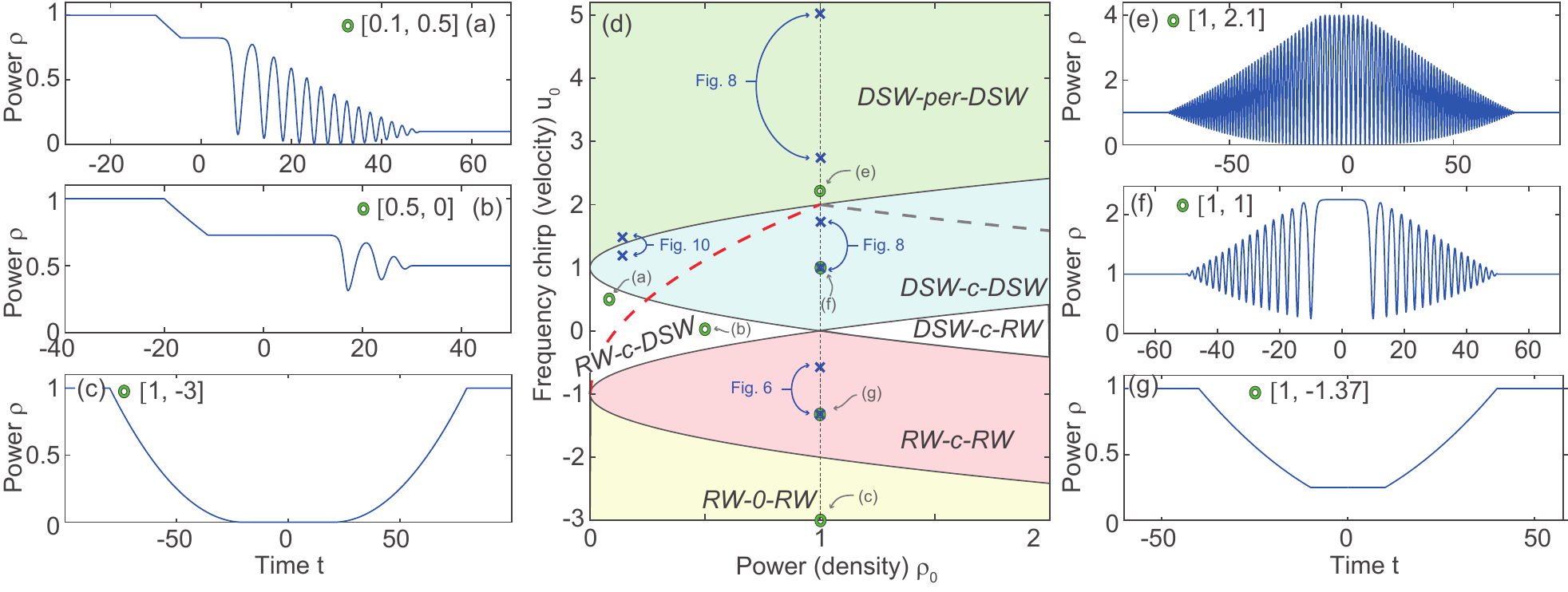}
     \caption{(d) Phase transition diagram for the Riemann problem of the NLSE: domains in the plane of input parameters $(\rho_0, u_0)$ of the step, giving rise to different wave-pair compositions of rarefaction (RW) and dispersive shock (DSW). The two waves in the pair can be connected through a constant state (-c-), zero density (vacuum state -0-); or an unmodulated nonlinear periodic wave (-per-). Above the red (gray) dashed lines, a vacuum is predicted to appear in the DSW-R (DSW-L). (a)-(c) and (e)-(g) are typical examples of output (normalized length $z_L=20$) from \red{Whitham modulation theory} marked with green circles \red{in panel (d)}. Experimental results are marked with blue crosses. 
     \ST{See also \cite{Supp} for a video illustrating the transitions between all the cases along the $(\rho_0=1)$ vertical line and for an extended version of this figure where, for completeness, we also report former experimental results from \cite{Xu17}.} 
}
\label{fig:parplane}
\end{figure*}

In the dispersive regime ruled by the NLSE, however, the solution of the same problem becomes more challenging since it requires to resort to Whitham modulation theory, which describes a potentially more rich dynamics. Following pioneering results \cite{Bikbaev95,El95} and our calculations outlined in Appendix B in \cite{Supp}, the result of such an approach is conveniently summarized in Fig. \ref{fig:parplane}, where we report in the plane $(\rho_0, u_0)$ the domains where different wave pairs are expected to emerge. Five different regimes are highlighted by domains of homogeneous color in Fig. \ref{fig:parplane}(d), whereas curves that separate the domains denote phase transitions among the different regimes. We also display in Figs. \ref{fig:parplane}(a-c) and (e-g), typical output power profiles of the different decay scenarios, as obtained from \red{Whitham modulation theory}.
The parameter values of such examples, as well as those of the experimental data are highlited in the parameter plane in Fig. \ref{fig:parplane}(d) by green circles and blue crosses, respectively.
 
The white regions in Fig. \ref{fig:parplane}(d) correspond to decay into a DSW and a RW connected by a constant power, which we indicate as DSW-c-RW. Experimental evidence for such scenario has been recently reported in fiber optics \cite{Xu17} and spin waves \cite{Janantha17}, for initial data lying on the horizontal line $u_0=0$. 
This indeed reproduces the well-known case study known as {\em shock tube problem} in gases or the equivalent dam breaking problem in hydrodynamics, i.e. the evolution of an initial step in density only. We also emphasize that the left ($\rho_0<1$) and right ($\rho_0>1$) white domains describe exactly the same physics, differing only for the direction of expansion of the DSW and RW pair. More precisely, the results in the semi-plane $\rho_0>1$ can be mapped in the semi-plane $\rho_0<1$ with the transposition $\rho_0 \rightarrow 1/\rho_0$ and $t \rightarrow -t$. 
\ST{
(for completeness, in \cite{Supp}, we reproduce Fig. 3 in Fig. S4(e) with additional results from Ref. \cite{Xu17}, which has allowed for a quantitatively accurate characterization of the cavitation which appears along the DSW when crossing above the dashed gray curve). 
}

In this paper, we are rather interested in the {\em piston problem}, that is a jump in velocity over a homogeneous density, which is described in Fig. \ref{fig:parplane}(d) by the vertical dotted line $\rho_0=1$. Typical \red{theoretical examples} (green circles) are shown in Figs. \ref{fig:parplane}(c,e,f,g), whereas the experimental results that will be presented in this paper are marked by blue crosses. 
Let us consider such line, starting from negative values $u_0=-|u_0|$. This corresponds to a negative (positive) equivalent velocity for $t<0$ ($t>0$), or, in other words, to initial velocity vectors pointing outward as shown in Fig. \ref{fig:cartoon}(d), which is equivalent to have two retracting pistons producing two symmetric RWs that expand in opposite directions. Such RWs can be quantitatively characterized by standard hydrodynamic theory applied to the dispersionless NLSE \cite{Leveque_book,Kevorkian_book} (in this case Whitham theory would give exactly the same result due to the smoothness of the solutions). As shown by the \red{snapshots} depicted in Figs. \ref{fig:parplane} (c) and (g), the two RWs smoothly connect the quiescent input state $\rho=\rho_0=1$ to a ``rarefied" state of lesser power $\rho_i=(1-|u_0|/2)^2$ and zero velocity $u_i=0$,
\ST{consistently with Eq. (\ref{rhoi})}.
The expansion of the right RW is determined by the velocities $\tau_1$ where its leading edge connects to the input state $\rho=1$ and $\tau_2$ of its trailing edge where it connects to the rarefied state $\rho_i$. We obtain for these velocities
\begin{equation} 
\tau_{1}=1 + |u_0|;  ~\tau_{2}=1 - \frac{|u_0|}{2},  \label{vel2RW}
\end{equation}
whereas the left RW expands with symmetric velocities are $\tau_{4}=-\tau_{1}$ and $\tau_{3}=-\tau_{2}$. 
Furthermore, when the two retracting pistons are fast enough, i.e. for $|u_0| >2$, the intermediate state becomes a zero density state or vacuum (see Fig. \ref{fig:parplane} (c)). 

When $|u_0|$ decreases the RWs become progressively shallower up to the limit $u_0 \rightarrow 0$ for which $\rho_i \rightarrow 1$ and the density remains obviously flat upon evolution.  
However, when $u_0$ crosses the line $u_0=0$ becoming positive, the behavior drastically changes because now the virtual pistons are pushing the fluid as in Fig. \ref{fig:cartoon}(b). As a result, two DSWs appear, being separated by a flat plateau with density $\rho_i$ which becomes, in this case, larger than the quiescent density $\rho=1$. This is the case shown in dimensionless form in the example in Fig. \ref{fig:parplane}(f) and discussed above in more detail with reference to Fig. \ref{fig:2shocks}. What is important to point out here is that, when $u_0$ increases, the amplitude of the oscillations in the DSW increases and the constant plateau (at fixed $z$) shrinks. At the threshold $u_0^{th}=2$, such amplitude reach its maximum and the DSWs start to cavitate (the bottom of the oscillations touches zero power), while the plateau shrinks to zero (i.e. the DSWs are glued back to back). Increasing $u_0$ further above this threshold, a phase transition to a new regime occurs. The new pattern features two DSWs connected by an unmodulated nonlinear periodic wave (DSW-per-DSW), as displayed in Fig. \ref{fig:parplane}(e) (more theoretical details are given in Appendix B in \cite{Supp}). This new regime, which was firstly pointed out by \red{Bikbaev} \cite{Bikbaev95}, is characteristic of the photon fluid and bear absolutely no similarity in gas dynamics. Importantly the transition from a DSW-c-DSW  to a DSW-per-DSW (cyan to green domain in Fig.\ref{fig:parplane}(d)) is not a prerogative of the vertical line $\rho_0=1$, which describes the canonical piston problem. Conversely, according to Whitham modulation theory, it can occur for generic Riemann step-initial data that contains also a concomitant jump in density ($\rho_0 \neq 1$), or, in other words, for a mixed shock-tube and piston problem. In this case, which yields non-symmetric DSWs, more generally the threshold reads as  \red{(for $\rho_0<1$)}
\begin{equation} 
u_0^{th}=1+ \sqrt{\rho_0},  \label{threshold}
\end{equation}
which correctly reduces to $u_0^{th}=2$ for the pure piston problem ($\rho_0 \rightarrow 1$). Since cavitation occurs above the red dashed curve in Fig. \ref{fig:parplane} (d), for the pure piston problem cavitation occurs exactly at threshold. Conversely, in the generic or mixed case, cavitation can occur below the threshold, as we will show in the experiment.


\section{\label{sec:exp} Experiment}
This section is devoted to describe our experimental technique implemented to investigate the piston-like dynamics, and observe the expected phase transitions associated with change of the initial condition in the dispersive Riemann problem.

\subsection{\label{sec:setup} A - Experimental setup}

\begin{figure}[htb]
\centering
\includegraphics[width=8.3cm]{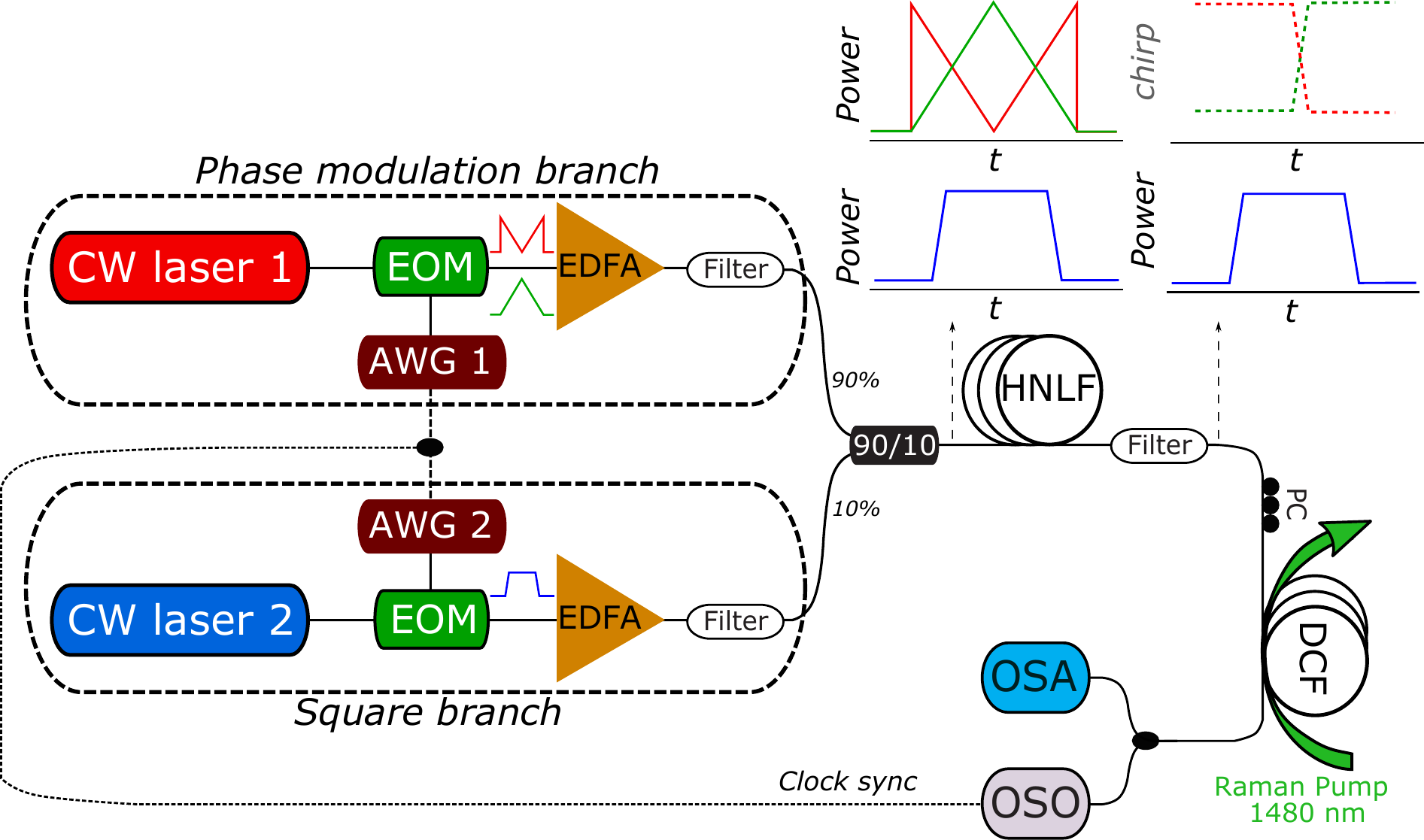}
\caption{Sketch of the experimental setup: EOM electro-optic modulator; EDFA, erbium-doped fiber amplifier; AWG, Arbitrary wave generator; OSO, optical sampling oscilloscope; OSA, Optical Spectrum Analyser; HNLF, highly nonlinear fiber, $L_H=500$ m; DCF, dispersion compensating fiber, $L=15$ km; PC, polarization controller. The choice of intensity modulation (M- or triangularly-shaped sketched in red and green, respectively) in the upper block (Phase modulation branch) is transformed into a stepwise chirp (either descending or ascending) of the beam at $\lambda_2$ at the input of the DCF.
}
\label{fig:f4}
\end{figure}

Our experiment takes advantage of a full fiber set-up which exploits state of the art telecommunication technology as sketched in Fig. \ref{fig:f4}. The setup has been specifically designed to face two key challenges: (i) to impress a \red{consistent and rapid} frequency modulation on the input signal; and \red{(ii) to compensate for the fiber loss in the main fiber (DCF).
The frequency modulation should be strong enough to produce} chirps up to $\sim \pm 10$ GHz, over relatively long duration (about 0.5 ns, which is long enough to observe the full development of DSW envelopes typically made of tens of modulation periods, as we will show below). 
\ST{Moreover, the transition to switch from negative to positive values of frequency and vice versa, must be ultra-fast so to approximate the ideally instantaneous stepwise variation.}
To this end we resort to an all-optical method based on cross-phase modulation (XPM) as in Ref. \cite{Varlot13} in order to be able to generate these large frequency chirps. This corresponds to a  maximum phase value of 10$\pi$, well above the typical $\pi$ rad., the characteristic maximum value accessible with standard phase modulators used in telecommunication applications. 

\begin{figure}[htb]
\centering
\includegraphics[width=8.3cm]{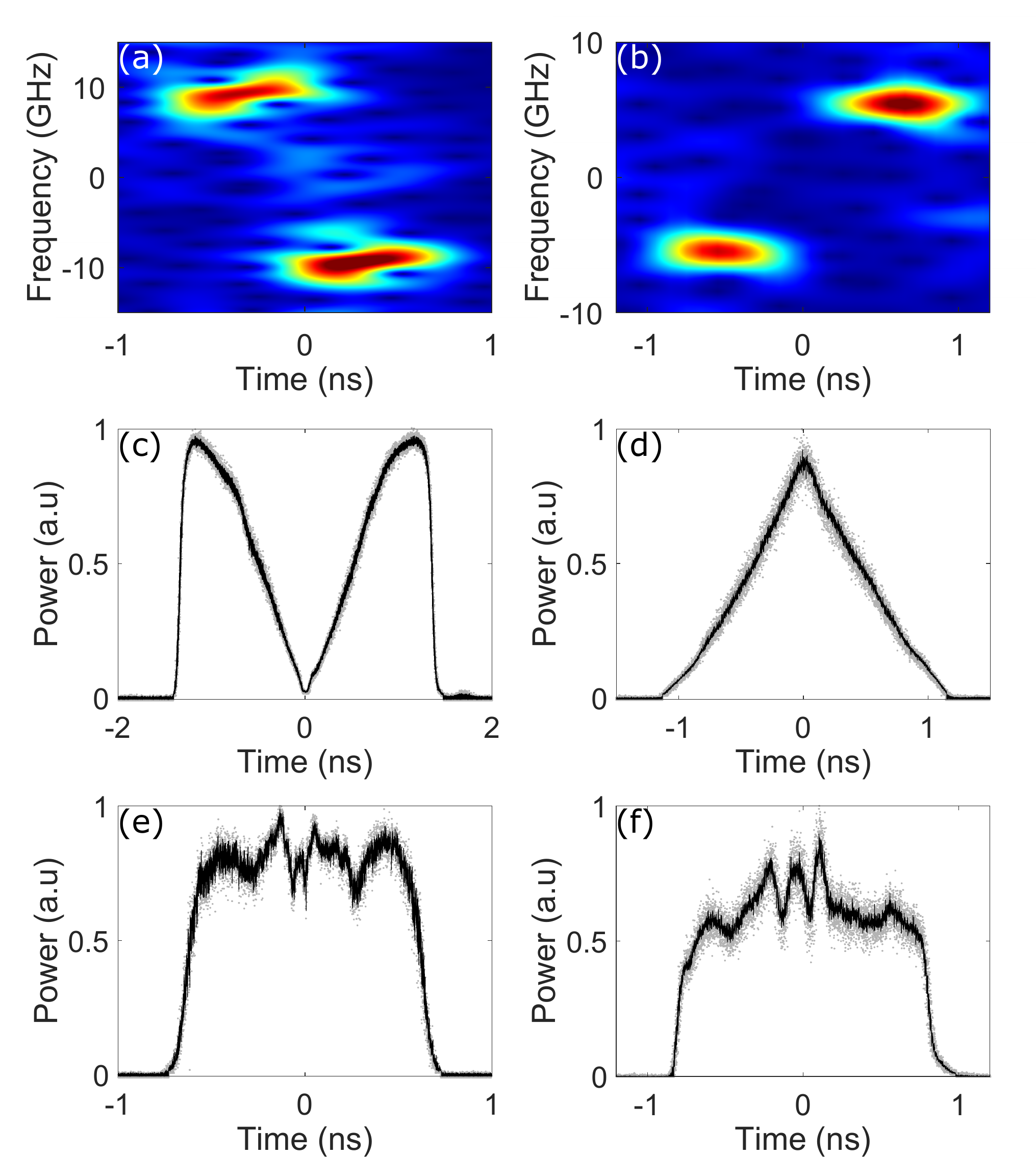}
     \caption{Experimental characterisation of the input to the DCF: (a,b) spectrograms $\Delta f$ vs time $T$ (dark red and blue stands for the maximum and minimum \red{power density} 
     respectively); (c,d) Temporal profile of the optical signal that induces all-optically the frequency modulation;
     (e,f) Power profile of the frequency modulated rectangular pulse. 
     Left column (a,c,e) and right column (b,d,f) refer to the case of \red{descending and ascending}  step-like frequency variation
     mimicking a pushing or retracting pair of pistons, respectively.
}
\label{fig:expinput}
\end{figure}

The main idea behind the preparation of the Riemann-like input (point (i)) is to use two laser sources at slightly different wavelengths suitably modulated in amplitude, which are combined and injected in a first fiber, namely a highly nonlinear fiber (HNLF).
The HNLF has the key role of transforming the amplitude modulation of one of the laser sources into the frequency modulation of the other source via XPM. The two lasers are independently modulated as sketched by the two main blocks denoted in Fig. \ref{fig:f4} as ``Phase Modulation Branch" and ``Square branch", respectively. In particular, in the lower block (Square branch) we make use of a continuous laser diode emitting at $\lambda_2=1561$ nm which is intensity modulated by an electro-optic modulator (EOM) driven by an arbitrary waveform generator (AWG2), amplified in an Erbium doped fiber amplifier (EDFA) and spectrally filtered to remove the amplified spontaneous emission in excess. The intensity modulation produces a train of square pulses with 25 MHz repetition rate. The pulses provide constant power (density) over $2$ ns duration, which is expected to be a sufficiently wide temporal window for the DSWs to develop in the piston-like experiment. 
Conversely, the upper block (Phase Modulation Branch) is devoted to impress a proper phase modulation to the beam at $\lambda_2$. To this end we start from a continuous laser diode emitting at $\lambda_1=1539$ nm, and impress an intensity modulation via a second EOM driven by AWG1, after which the signal is again amplified and filtered. The modulation is synchronous with that of the other block, whereas the intensity waveform can be chosen to be either M-shaped (sketched in red in Fig. \ref{fig:f4}; see also Fig. \ref{fig:expinput}(c) for its real profile) or triangularly shaped (sketched in green in Fig. \ref{fig:f4}; real profile in Fig. \ref{fig:expinput}(d)). 
The output of the two blocks are combined through a 90:10 coupler, so that we launch in the HNLF a dual wavelength signal constituted by square pulses with flat phase with typical peak power $P_2=40$ mW superimposed  to a more powerful beam (peak $P_1(T)_{max}=7$ W) of different color and suitable temporal shape (either M-like or triangular). The linear polarization of the two beams is controlled to be parallel in order to maximize the effect of XPM. During propagation in the HNLF, the beam at $\lambda_1$ induces via XPM an output phase modulation $\phi_2(t)=2 \gamma_{H} L_{H} P_1(T)$ over the beam at $\lambda_2$, $\gamma_{H}= 12$ (W.km)$^{-1}$ and $L_{H}=500$ m being the nonlinear Kerr coefficient and the length of the HNLF, respectively. This corresponds to a chirp $\Delta \omega(T) = -d\phi_2(T)/dt= -2 \gamma_{H} L_{H} dP_1(T)/dT$, and in turn to a profile of the equivalent initial gas velocity $u(t)=\Delta \omega(T) T_0=2\pi \Delta f(T) \sqrt{k''/(\gamma P_0)}$. The abrupt change of slope in the M-shaped or triangular intensity modulation is converted, due to the derivative that links the frequency to the phase, into a step-like variation of the chirp or equivalent gas velocity, which is expected to be either descending (for the M-shaped case) or ascending (for the triangular shape), as sketched in Fig. \ref{fig:f4}.
At the output of the HNLF the beam at $\lambda_1$, as well as the multiple sidebands produced by four-wave mixing of the input beating between $\lambda_1$ and $\lambda_2$,  are filtered out through a bandpass filter. The remaining, strongly chirped beam at $\lambda_2$ constitutes the Riemann-like input which is injected in the main fiber, i.e. the $L=15$ km long DCF. In such fiber it becomes crucial to compensate the losses (point (ii)) which amounts to 0.5 dB/km or nearly $80 \%$ total loss. This is performed by exploiting the Raman gain from a counterpropagating pump at $\lambda_2=1480$ nm (for more details see Ref. \cite{Xu17}). Finally, the output of the DCF is monitored both spectrally, by means of an optical spectrum analyzer (OSA) and in time domain by means of an optical sampling oscilloscope (OSO) synchronized by the clock of the two AWGs.

In Fig. \ref{fig:expinput} we display the results of the experimental characterisation of the input to the DCF.
The case of descending step-like variation produced via the M-shaped modulation is shown in the left column of Fig. \ref{fig:expinput} (pushing piston in Fig. \ref{fig:cartoon}).
In particular, Fig. \ref{fig:expinput}(a) shows the measured spectrogram, i.e. the frequency deviation $\Delta f=\Delta \omega/2\pi$ from the input carrier frequency as a function of time $T$ (see \cite{Supp} for further details).
The beam clearly exhibits an abrupt jump in frequency around $T=0$, with chirp excursion between $\pm \Delta f_0$ with $\Delta f_0=9.4$ GHz, and an estimated rise time (10 to 90$\%$) $T_r=50$ ps.
Being symmetric around $\Delta f=0$, this frequency modulation realises the step-like variation from $u_0$ to $-u_0$ formerly introduced in Figs. \ref{fig:cartoon}-\ref{fig:parplane}, with $u_0=2\pi \Delta f_0 \sqrt{k''/(\gamma P_0)}$. 
The actual trace of the intensity-modulated optical signal that drives the frequency modulation process via XPM is shown in Fig. \ref{fig:expinput}(c), obtained at the input of the HNFL.
It is evident that such driving M-shaped signal is characterized by a sharp change of slope around $T=0$, while it exhibits uniform slopes over a large temporal window of $\sim 2$ ns, which exceeds the duration of the square pulses shown in Fig. \ref{fig:expinput}(e). The control of the linearity of the slopes is crucial to have flat frequency states in the input to the DCF as clearly observed in the spectrogram in Fig. \ref{fig:expinput}(a). Although this demonstrates that the process of all-optical modulation works in nearly ideal way for what concerns the desired frequency modulation, \ST{
it is important to mention that we had to implement a pre-compensation algorithm to optimize the shape of the electric field driving the modulator in order to obtain a quasi-flat pulse \cite{Supp,Braud15} as shown in Fig. \ref{fig:expinput}(e). Without this scheme, strong distortions due to Raman amplification and four-wave mixing between the pulses would not allow to achieve the results shown below.
}
The same type of measurements performed for the choice of triangular driving signal is reported in the right column of Fig. \ref{fig:expinput} (pulling piston in Fig. \ref{fig:cartoon}).
The spectrogram in Fig. \ref{fig:expinput}(b) clearly show the ascending nature of the frequency jump. In this case, however, in order to have tolerable distortion of the plateau of the square pulses, the achievable frequency jump is limited to a lower excursion $\Delta f_0 = 5.4$ GHz. In fact, the triangular shape implies, at variance with the M-shape, that the change of slope occurs at the maximum power of the signal trace. Larger jumps would require higher slopes and hence larger peak power of the triangular shape, which ends up causing much stronger distortions of the power plateau of the square pulses compared with that shown in Fig. \ref{fig:expinput}(f).

\subsection{\label{sec:results} B - Results}
\begin{figure}[htb]
\centering
\includegraphics[width=8.3cm]{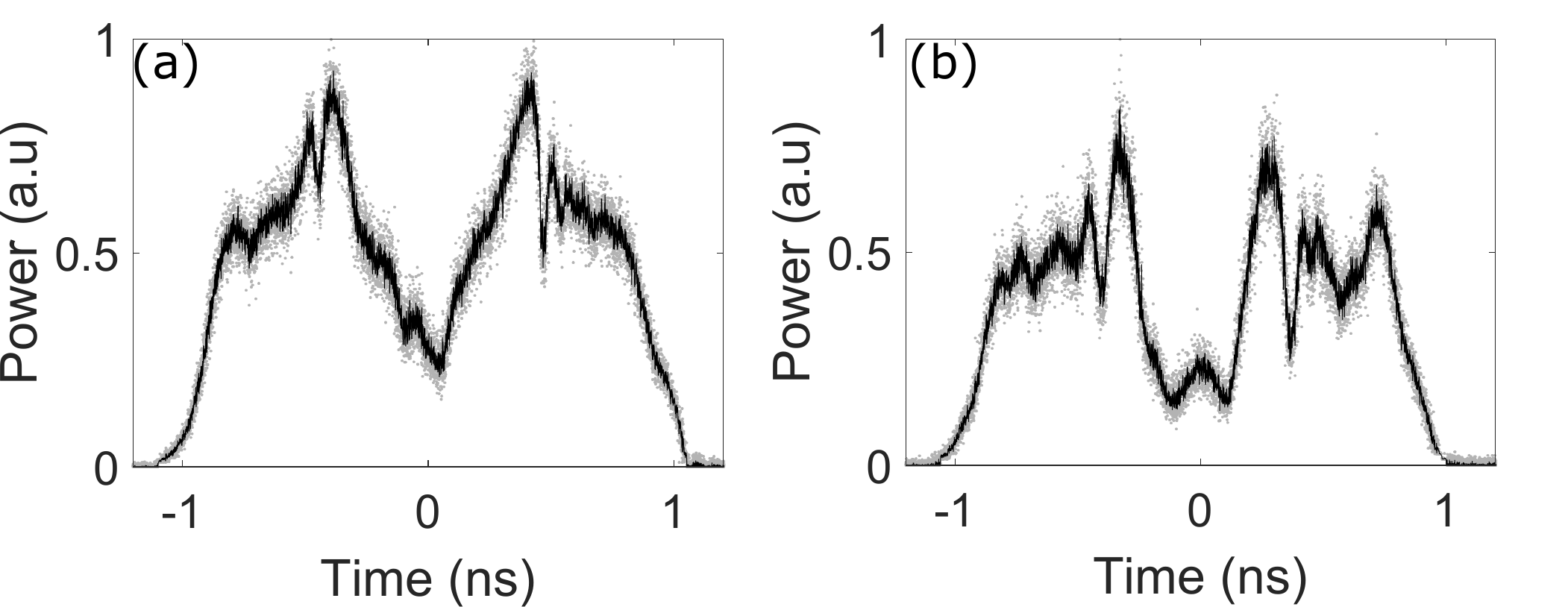}
     \caption{
     Rarefaction wave pair produced by an increasing input jump in frequency with $\pm \Delta f_0=5.4$ GHz, and power:
     (a) $P_0=165$ mW ($u_0=-0.63$); (b) $P_0=35$ mW ($u_0=-1.37$).
}
\label{fig:expA}
\end{figure}

\begin{figure}[htb]
\centering
\includegraphics[width=8.3cm]{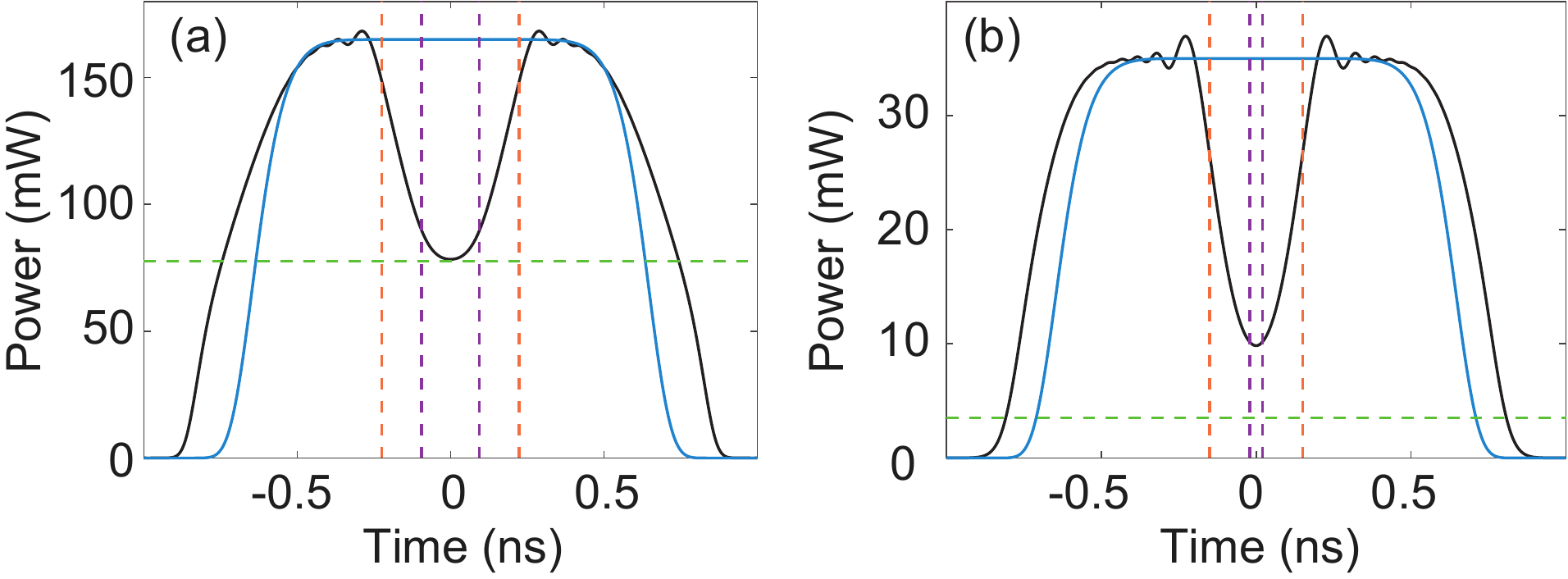}
     \caption{Simulations based on the NLSE corresponding to Fig. \ref{fig:expA}:
     (a) $P_0=165$ mW; (b) $P_0=35$ mW.
     The dashed vertical lines stand for hydrodynamic velocities calculated from Eq. (\ref{vel2RW}).
     \ST{The horizontal green line stands for $\rho_i P_0$, with $\rho_i$ from Eq. (\ref{rhoi})}.
     Here small numerical oscillations are due to Gibbs phenomenon \cite{Gibbs}.
}
\label{fig:simA}
\end{figure}

\begin{figure}[htb]

\centering
\includegraphics[width=8.3cm]{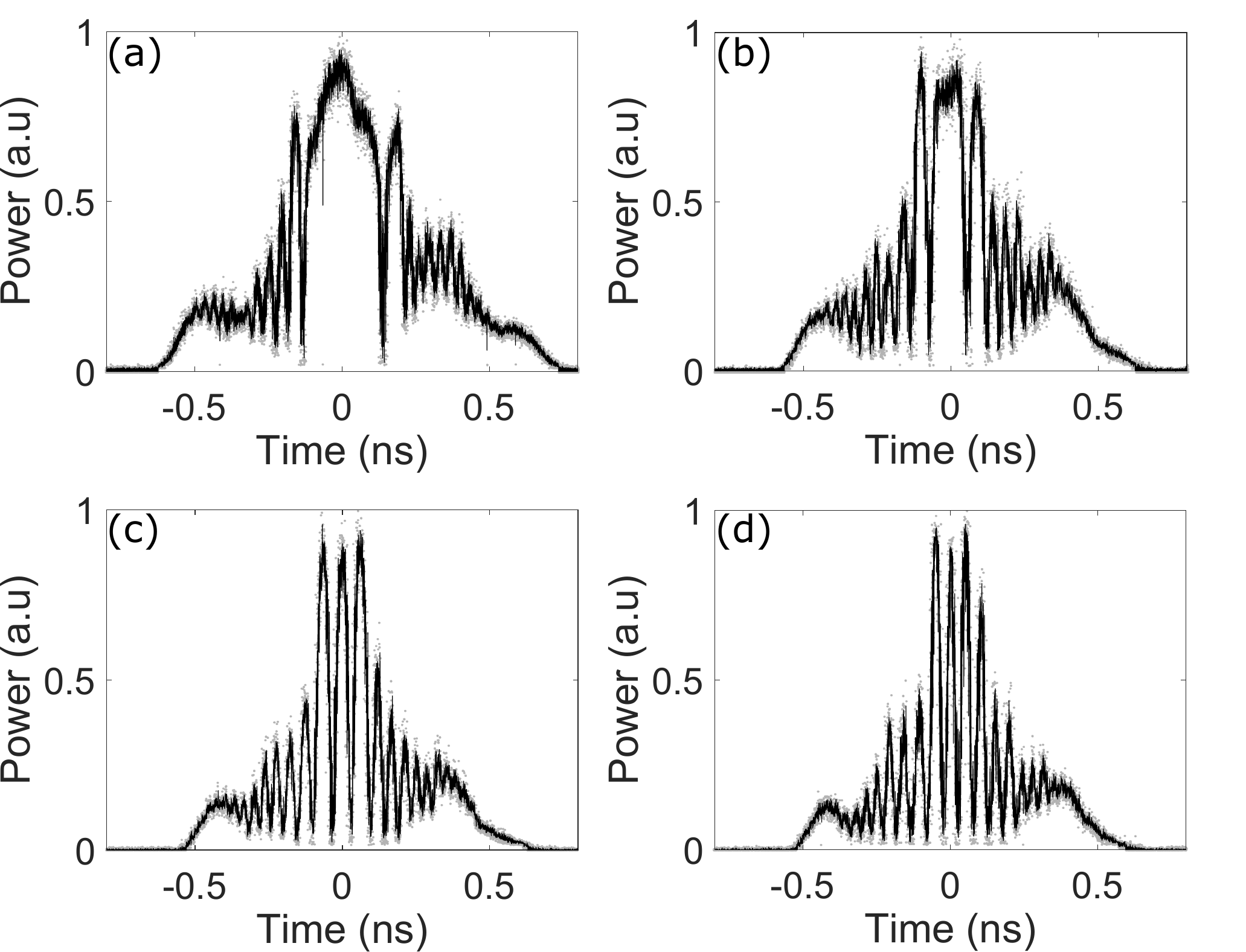}
     \caption{
Shock formation: DSW-c-DSW (a,b) and their phase transition to DSW-per-DSW (c,d), observed for $\Delta f_0=9.4$ GHz, and power:  
(a) $P_0=220$ mW ($u_0=0.95$); (b) $P_0=72$ mW ($u_0=1.66$); 
(c) $P_0=27$ mW ($u_0=2.71$); (d) $P_0=8$ mW ($u_0=4.97$).
}
\label{fig:expM}
\end{figure}

\begin{figure}[htb]
\centering
\includegraphics[width=8.3cm]{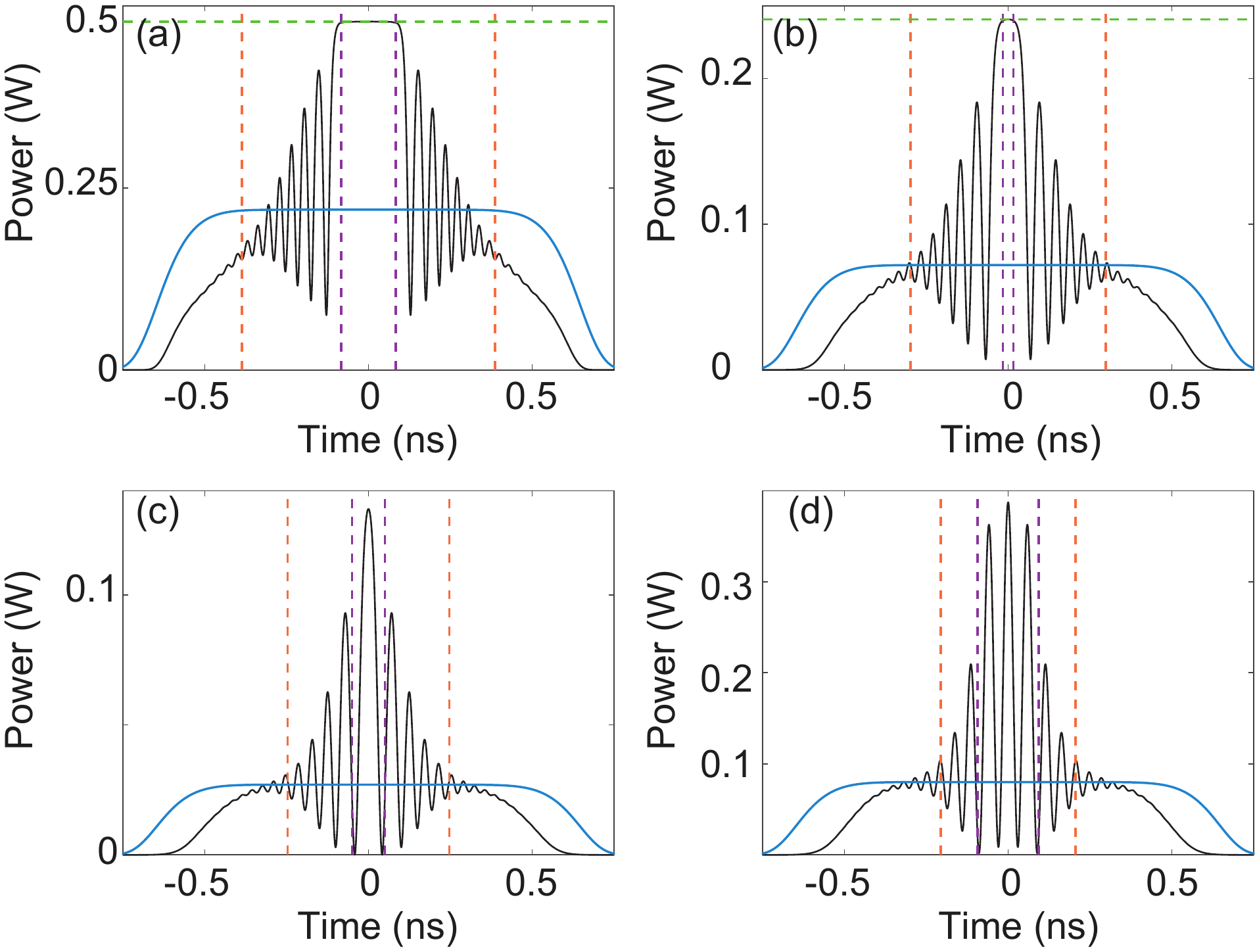} 
     \caption{Output profiles obtained from numerical simulation of the NLSE (\ref{nls}) corresponding to the results in Fig. \ref{fig:expM}:
     (a) $P_0=220$ mW; (b) $P_0=72$ mW; (c) $P_0=27$ mW; (d) $P_0=8$ mW.
     Vertical dashed lines arise from modulation theory:  linear DSW edges (orange) and soliton edges in (a,b) or periodic wave edges in (c,d) (magenta).
     In (a,b) \ST{the horizontal green line stands for $\rho_i P_0$, with $\rho_i$ from Eq. (\ref{rhoi})}.
}
\label{fig:simM}
\end{figure}

\begin{figure}[htb]
\centering
\includegraphics[width=8.3cm]{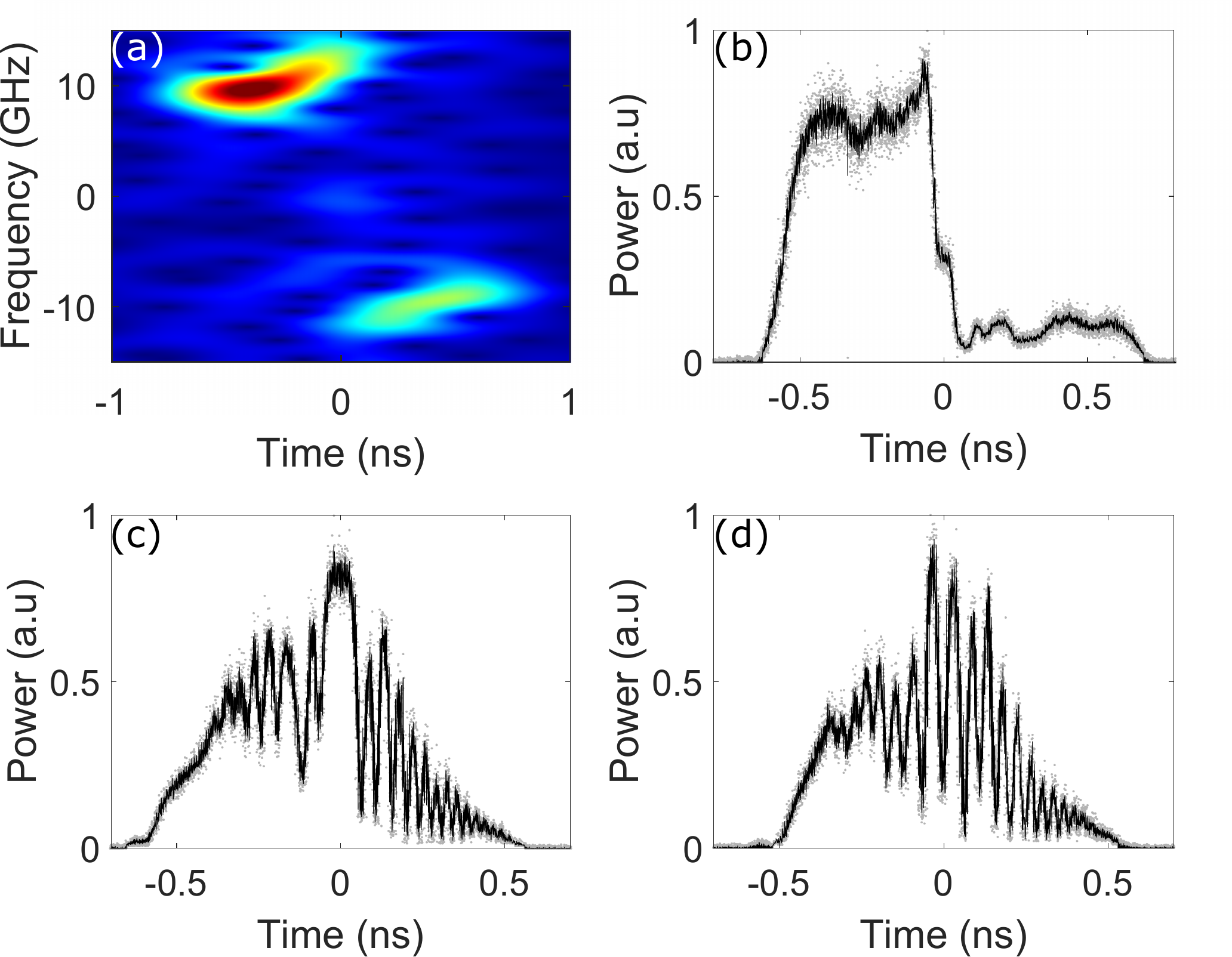}
     \caption{
     Transition from DSW-c-DSW to DSW-per-DSW in the asymmetric case, with fixed $\Delta f=9.7$ GHz and additional jump in power with nominal extinction ratio $P_R/P_L=0.15$:
     (a) spectrogram of the input; (b) profile of input power jump; 
     (c,d) output power profiles: (c) $P_0=240$ mW, below threshold;  (d) $P_0=90$ mW, above threshold.
}
\label{fig:expAsym}
\end{figure}

Here we discuss the outcome of the experiment which is summarized in Figs. \ref{fig:expA}-\ref{fig:expAsym}. The focus is to explore the transitions between the variety of generated wave pairs from an input with constant power ($\rho_0=1$) and step-like frequency variation, resembling the action of two counter-propagating pistons on constant density. Specifically, we discuss the results in order of increasing values of the equivalent velocity $u_0$, starting from negative values. 
Then we also address the shock transition for a more general Riemann input, namely a joint step-like variation of chirp and power (case $\rho_0 \neq 1$). Importantly, in all regimes, we find that the most stable and repeatable configuration is to operate at the fixed maximum achievable $\Delta f_0$ (i.e., the values illustrated in Fig. \ref{fig:expinput}(a,b) or Fig. \ref{fig:expAsym}(a)) and tune the effective velocity of the piston $u_0$ by changing the power of the modulated square pulse in input to the DCF, recalling that $u_0$ scales like $u_0 \propto \Delta f_0/\sqrt{P_0}$. 

Let us start with the case of negative $u_0$, which corresponds to the ascending step obtained with the triangular shape of the driving signal,  resembling the situation sketched in Fig. \ref{fig:cartoon}(e) and leading to the RW-c-RW case (pink area in Fig. \ref{fig:parplane}(d)).
We show in Fig. \ref{fig:expA} two typical output power profiles measured by means of the sampling oscilloscope and obtained for fixed $\Delta f_0 = 5.4$ GHz and two power levels $P_0=165$ mW and $P_0=35$ mW. 
The \red{blue crosses} in Fig. \ref{fig:parplane} highlights the location of the corresponding values of equivalent velocity, $u_0=-0.63$ and $u_0=-1.37$, respectively, in the domain of parameter plane featuring RWs.
The output power profiles in Fig. \ref{fig:expA} clearly show the formation in the center of the nearly square pulse of a wide hole or dark region.
The smooth edges of this hole constitute two optical RWs, that are driven by the initial condition that acts like a pair of retracting pistons.
In Fig. \ref{fig:simA} we show the corresponding output profiles obtained from numerical integration of the full NLSE (\ref{nls}), using a nearly ideal waveform, namely a super-Gaussian pulse profile (see blue curves in the figure)
with frequency modulation $\Delta \omega (T)=-u_0 T_0 \tanh (T/W_0)$, $W_0=20$ psec. The numerics shows a good qualitative agreement with the experiment, though the experimental trace clearly show a distortion, which is due to the non-perfectly flat plateau of the injected modulated pulse (see Fig. \ref{fig:expinput}(f)). Clearly, the hole is progressively dug during propagation until, in the case at  power $P_0=165$ mW, the bottom of the RWs touches on the intermediate or rarefied state at constant power $\rho_i P_0$ predicted by the dispersionless NLSE  (see dashed green horizontal line in Fig. \ref{fig:simA}(a)).
Conversely, at lower power $P_0=35$ mW, which corresponds to more negative $u_0$, while the intermediate state becomes considerably lower (dashed green in Fig. \ref{fig:simA}(b)),
both the experiment and the simulation exhibit only a slightly darker hole compared with previous case. This is due to the fact that, lowering the power results also in a slower dynamics and a shorter effective length ($z_L=L \gamma P_0$). 
As a result, the two RWs are expected to finally dig to the constant rarefied state at power $\rho_i P_0$ only at distances which far exceed the actual fiber length $L=15$ Km. For the same reason, the state RW-0-RW (yellow region in Fig. \ref{fig:parplane}) characterized by a bottom state which becomes a black (zero intensity) state remains elusive with our present setup, since it would need to further decrease $u_0$ below $u_0 = -2$, without, however, decreasing too much the effective length $z_L$.

Conversely, when $u_0$ increases towards zero, the RWs become shallower until, above the threshold $u_0=0$ (i.e., for positive $u_0$), a transition to the state DSW-c-DSW occurs (cyan area in Fig. \ref{fig:parplane}(d)).
In our experiment, such transition can be observed by reversing the step in frequency to become descending as in Fig. \ref{fig:expinput}(a). 
The relative output traces obtained for fixed $\Delta f_0 = 9.4$ GHz and two different input powers, $P_0=220$ mW and $P_0=72$ mW, corresponding to $u_0=0.95$ and $u_0=1.66$ (blue \red{crosses}) in Fig. \ref{fig:parplane}), are shown in Fig. \ref{fig:expM}(a) and (b), respectively. We clearly observe the formation of two nearly symmetric DSWs connected by a constant intermediate state, which, in contrast with previous case, marks the highest power of the waveform.
The experimental trace can be compared with the corresponding simulations of the full NLSE (\ref{nls}) reported in Fig. \ref{fig:simM}(a,b), and performed with ideal input (modulated \red{super-Gaussan} pulses).  
A good qualitative agreement is obtained also in this regime, with discrepancies arising mainly from the imperfect flat plateau of the injected real pulses shown in Fig. \ref{fig:expinput}(e).
We also point out that the location of the linear and soliton edges of the DSWs exhibit a satisfactory agreement with the predictions of Whitham theory (Eqs. (\ref{Whitham2DSW}), vertical dashed lines in Fig. \ref{fig:simM}(a,b)), 
which considerably extend the applicability of such theory beyond its nominal asymptotic limit of validity.
Noteworthy, comparing Fig. \ref{fig:expM}(a) with Fig. \ref{fig:expM}(b), as well as \ref{fig:simM}(a) with Fig. \ref{fig:simM}(b) for the numerics, we notice that the intermediate state that connects the DSWs shrink when $u_0$ increases,
(i.e., when power decreases). At the same time, the DSWs increase their contrast, exhibiting lower minima and higher maxima. When the intermediate state shrinks to zero the state DSW-c-DSW is no longer sustainable.
According to Whitham theory this occurs at threshold $u_0^{th}=2$, above which the two DSWs are connected through a periodic nonlinear wave (see Appendix B in \cite{Supp} for more technical details). 
We have checked quantitatively that this phase transition, which possess no analogy in the realm of classical non-dispersive fluids, can be observed by decreasing further the power in order to increase $u_0$ above threshold.
The output traces relative to $P_0=27$ mW and $P_0=8$ mW ($u_0=2.71$ and $u_0=4.97$, respectively, \red{blue crosses} in Fig. \ref{fig:parplane}) are reported in Fig \ref{fig:expM}(c,d).
They clearly show the DSWs to be connected through a periodic wave instead of a constant state, in good agreement with the corresponding simulations displayed in Fig. \ref{fig:simM}(c,d).
Note that, in this case, the DSWs no longer possess soliton edges, but rather connect smoothly to the periodic wave at temporal locations that can be calculated by Whitham modulation theory 
(magenta dashed lines in Fig. \ref{fig:simM}(c,d); technical details in Appendix B in \cite{Supp}).

Finally, we have also tested the peculiar transition DSW-c-DSW to DSW-per-DSW (cyan and green areas in Fig. \ref{fig:parplane} (d)) in the more general case where the Riemann initial data involve a jump both in frequency and power . We refer to this regime as the {\em asymmetric case} since this type of initial condition breaks the symmetry between the left-going and right-going DSWs. In terms of analogy with gas dynamics this initial datum would correspond to two pistons pushing on two sectors with gases at different density, i.e. a mixed case where two canonical initial conditions of the piston type (pure jump in velocity) and shock tube type (pure jump in density) are combined together. It is important to emphasise that this mixed type of initial conditions are difficult to realize in any common facility for standard fluids, such as gas dynamics tube experiments or shallow water tanks, and we are not aware of experimental results obtained for such case. Conversely, the flexibility of our setup allows us to easily access such more general type of Riemann input. Indeed in our setup, we can act on the square wave branch, and in particular on the AWG2, to impress an additional jump in power in the middle of the square pulse, leaving unaltered the branch in the setup that is responsible for the step-like frequency variation.
The result of the experimental characterization of the input is shown in Fig. \ref{fig:expAsym}(a,b). The spectrogram in Fig. \ref{fig:expAsym}(a) clearly indicates that the step-like variation of frequency chirp is preserved, with a symmetric jump between $\Delta f_0$ and $-\Delta f_0$, with $\Delta f_0=9.7$ GHz. In this case, however, it is clear that the positively and negatively chirped portions have strongly different intensity. This is further clear from the associated power profile of the input to the DCF which is displayed in Fig. \ref{fig:expAsym}(b).
As shown, we illustrate the regime characterized by a a small extinction ratio $\rho_0=P_R/P_L=0.15$ or, in other words, by a large difference in power between the left and right state. This case is indeed the most interesting one because the right DSW is expected to exhibit always a cavitation (or vacuum) point in the whole region of existence of the DSW-c-DSW, at variance with the symmetric case where cavitation marks exactly the transition point $u_0=2$. Indeed, whenever $\rho_0<0.25$, one can see from Fig. \ref{fig:parplane} that the whole domain of existence of DSW-c-DSW, as well as that of DSW-per-DSW, lies above the dashed red line which stands for the threshold of appearance of a vacuum point (see also Appendix B in \cite{Supp}).
\ST{The threshold for the transition DSW-c-DSW to DSW-per-DSW occurs for fixed $\rho_0$ at the critical value $u_0^{th}$ given by Eq. (\ref{threshold}).
}
 
Figure \ref{fig:expAsym}(c,d) report the temporal power profiles observed at the output of the DCF, when operating at constant jump in frequency ($\pm 9.7$ GHz), constant extinction ratio $\rho_0=0.15$, and a variable power $P_0$ in order to vary the normalized chirp $u_0$.  At $P_0=240$ mW, which corresponds to a normalized chirp 
$u_0=0.94<u_0^{th}=1.39$
the evolution is expected to give rise to DSW-c-DSW. This is shown in Fig. \ref{fig:expAsym}(c), where we clearly observe the constant state (plateau) separating two strongly asymmetric DSWs. In particular the right DSW exhibits a cavitation point falling within its envelope.
When the power is decreased to $P_0=90$ mW, which correspond to the above threshold value 
$u_0=1.52>u_0^{th}$
the constant state disappears and the two DSWs appear to be connected by a periodic wave, as shown in Fig. \ref{fig:expAsym}(d).

\begin{figure}[htb]
\centering
\includegraphics[width=8.3cm]{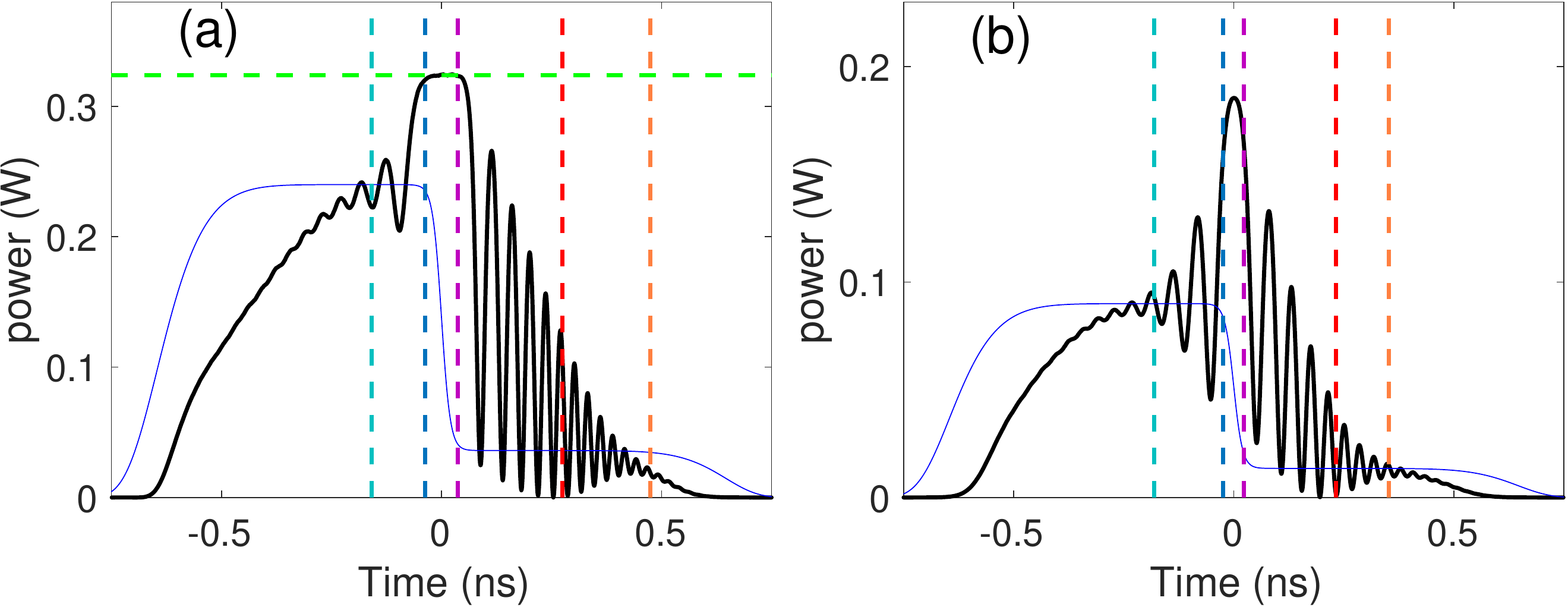} 
    \caption{Output profiles obtained from numerical simulation of the NLSE (\ref{nls}) corresponding to the results in Fig. \ref{fig:expAsym}(c,d):
    (a) $P_0=240$ mW; (b) $P_0=90$ mW. Here $\rho_0=P_R/P_L=0.15$. The vertical dashed lines stand for the vacuum point of the R-DSW (red),
    and the edge velocities of the DSWs (orange and cyan for linear edges, blue and purple for soliton or inner edges).   
    In (a) the horizontal green dashed line stands for $\rho_i$.
}
\label{fig:numAsym}
\end{figure}

In Fig. \ref{fig:numAsym}, we report the corresponding NLSE simulations with ideal initial data (blue curves in the figure). The agreement with the observed profiles is reasonably good, given the fact that the input profile exhibits considerable deviation from the ideal case, as \red{shown} in Fig. \ref{fig:expAsym}(b).
Importantly note that Whitham theory allows to predict with extremely good accuracy the location of the vacuum point in both regimes. \ST{Conversely, since the DSWs become strongly asymmetric, modulation theory allows to predict with good accuracy the temporal location of the edges (see orange and purple dashed vertical lines in Fig. \ref{fig:numAsym}) of the DSW-R which is quite extended, whereas it is less accurate for the DSW-L, which is substantially narrower.}

\section{\label{sec:end} Conclusions}
\ST{
In summary, we have fully characterized the phase transitions associated with the Riemann problem in a local fluid of light whose behavior is ruled by the universal NLSE. The physical piston is replaced by a stepwise optical pulse over which a quasi-instantaneous frequency chirp is imprinted, allowing to reproduce any velocity-density pair input conditions which mimic problems that span from the pure piston problem to its mix with the shock tube problem. These specially designed envelope pulses are launched in a defocusing optical fiber \red{whose} losses are actively compensated. In this way, the system is modeled by \red{integrable} NLSE and quantitative comparisons with theoretical developments from the Whitham modulation theory can directly be performed with experiments. We have been able to report (i) a comprehensive study of the phase transitions that occur in the dispersive piston problem ruled by the defocusing NLSE; (ii) the first observation of a new regime, which has no similarity in gas dynamics, featuring two DSWs connected through an unmodulated periodic wave; (iii) the observation of asymmetric DSWs and their critical transition to the fully undulatory solution that follows from the most general Riemann problem involving a simultaneous jump both in power and chirp.  All these observations are in very good agreement with theoretical predictions and numerics. This confirms that these \red{transparent} fiber-based optical systems are peerless testbeds to investigate the extension of gas dynamics problems to superfluid regimes by taking benefit of the analogy between optics and fluid-dynamics supported by the universality of the NLSE. 
}

\section{Acknowledgments}
The present research was supported by IRCICA (USR 3380 CNRS, projet IRCICA 2020), Agence Nationale de la Recherche (Programme Investissements d’Avenir); Ministry of Higher Education and Research; Hauts de France Council; European Regional Development Fund (Photonics for Society P4S, FUHNKC, EXAT). The authors are grateful to L. Bigot, E. Andresen and IRCICA-TEKTRONIX European Optical and Wireless Innovation Laboratory for technical support about the electronic devices. Discussions with A. Kamchatnov, N. Pavlov and P. Sriftgizer are gratefully acknowledged. The authors thanks T. Sylvestre for providing the HNLF fiber.


\begin{thebibliography}{99}
\bibitem{warnvelocity} we refer to $\tau=t/z$ as a velocity, as commonly done in shock wave theory, though it corresponds dimensionally to an inverse velocity.
\bibitem{stepnormalization} Non symmetric jump in $u$ only introduces a net drift in the problem, whereas the step in $\rho$ can always be normalized to have $\rho=1$ over one of the boundaries, that we choose to be the left side.
\end{thebibliography}
\end{document}